\documentclass[aps,prb,twocolumn,showpacs,a4paper]{revtex4-1}
\usepackage{amsfonts}
\usepackage{amsmath}
\usepackage{dcolumn}
\usepackage[english]{babel}
\usepackage{graphicx}
\usepackage[small,bf,hang,raggedright,FIGTOPCAP]{subfigure}
\usepackage{varioref}
\input epsf
\usepackage{epstopdf}
\begin{document}
\title{Effect of hydrogen on the microstructure evolution of nanocrystalline silicon}
\author{Giorgia Fugallo$^1$, Alessandro Mattoni$^2$, Luciano Colombo$^3$}
\email[e-mail: ]{luciano.colombo@dsf.unica.it}
\affiliation{
$^1$ Physics Department, King's College, London, Strand, London WC2R 2LS, United Kingdom\\
$^2$ Istituto Officina dei Materiali del CNR (CNR-IOM), c/o Dipartimento di Fisica, Cittadella Universitaria, I-09042 Monserrato (CA), Italy\\
$^3$ Dipartimento di Fisica, Universit\`a di Cagliari, Cittadella Universitaria, I-09042 Monserrato (Ca), Italy
}
\date{\today}
\begin{abstract}
We investigated the effect of dissolved hydrogen on the microstructure evolution of nanocrystalline silicon. Through molecular dynamics simulations we characterize the local and overall structural features of several hydrogenated samples by the quantitative calculation of their crystallinity. We prove that hydrogen incorporation is detrimental for the recrystallization and accumulates at grain boundaries, thus increasing their effective area. We further observe the formation of silicon hydrogen complexes at very high hydrogen concentration.
\pacs{64.70.Nd, 61.46.Hk, 81.10.Aj, 07.05.Tp}
\end{abstract}
\maketitle
Nanocrystalline silicon (nc-Si) belongs to the wide class of mixed-phase amorphous/crystalline systems of large technological 
impact.\cite{spinella,cleri-nc,pizzini,kuntz,awaji,nanodiamond,zhang}
In particular, nc-Si has recently attracted considerable interest as an efficient and comparatively low cost material for third generation solar cells.\cite{conibeer} Indeed, the coexistence in the same texture of amorphous and crystalline regions optimally combines a relatively high conductivity with a relatively high optical absorption. 

Among many other processing issues, hydrogenation (whether intentional or due to contamination) is critical in affecting the electronic properties of nc-Si. Typically, hydrogen is inserted in order to passivate dangling bonds in the amorphous phase, thus restoring a proper energy gap and reducing the concentration of recombination centers. This, in turn, enables applications of hydrogenated nc-Si (nc-Si:H) in optoelectronics and photovoltaics through an increased conductivity of photo-excited carriers. In addition, it has been recently demonstrated that H affects the nature of the electronic localization in the a-Si phase of nc-Si samples and promotes quantum confinement effects.\cite{bagolini-prl} Typical values of the hydrogen content in real samples is around 5\%,\cite{tauc} but locally higher values are indeed possible.

Hydrogenation is likely to affect the atomic scale morphology and microstucture evolution of nc-Si systems as well. In particular, it strongly affects the recrystallization kinetics, i.e. the spontaneous transformation from the amorphous to the crystalline phase.\cite{olson} In fact, H atoms can change the relative stability of the two phases and the mobility of the phase boundaries. For example, it is known that the mobility of a planar amorphous-crystalline (a-c) silicon interface depends on the localization of hydrogen impurities at the boundary where coordination defects are abundant.\cite{olson}  In the case of nc-Si materials  the physics of the recrystallization is further complicated because of the morphological complexity introduced by the network of phase boundaries.

Although the above qualitative picture is commonly accepted, its atomic-scale understanding is still largely missing. In order to clarify this issue we performed molecular dynamics (MD) simulations addressed to identify the role of hydrogen on the nc-Si recrystallization phenomena. The theoretical study of this topic is difficult because it requires an accurate description of any relevant atomic scale feature (like, for instance, the local degree of structural order or the hydrogen accumulation), while including a realistic description of the large scale a-c morphology (i.e. the grain size distribution, as well as the network of a-c interfaces). To this aim, we have taken profit of recent advances in computer modeling of nc-Si and the development of quantitative analysis tool (like, for instance, the local crystallinity).\cite{mattoni-prl-nc,mattoni-prb-nc,apl} By the present atomistic simulations we provide evidence that hydrogen at high concentration increases the disorder of the system and it lowers the recrystallization kinetics.

Because of the above interplay between atomic- and large-scale features, special care was played in generating our computational samples. We first obtained a set of pure nc-Si systems with different crystal-to-amorphous ratios. To this aim, by quenching from the melt, we generated a bulk a-Si sample within a periodically repeated cell of dimensions $L_x=2.5$ nm,  $L_y=25$ nm, and $L_z=1.25$ nm, containing as many as 10$^5$ Si atoms. A distribution of isolated grains was then inserted as in Ref.[\onlinecite{mattoni-prl-nc}]. The grains were extracted from a perfect silicon lattice in such a way that their $[100]$ crystallographic direction was parallel to the $x$ direction of the simulation cell. Each grain was then randomly rotated around $x$ in order to emulate the experimental regime of casual nucleation. The system was finally thermally annealed at 1200 K for as long as $3$ ns. The interatomic forces were calculated according to the environment dependent interatomic potential\cite{edip-a-si,edip-spe} and the equations of motions were integrated by a velocity-Verlet algorithm with a time step of 1.0 fs. Temperature was controlled by a rescaling atomic velocities.

\begin{figure}
\includegraphics[width= 0.38\textwidth, angle=0]{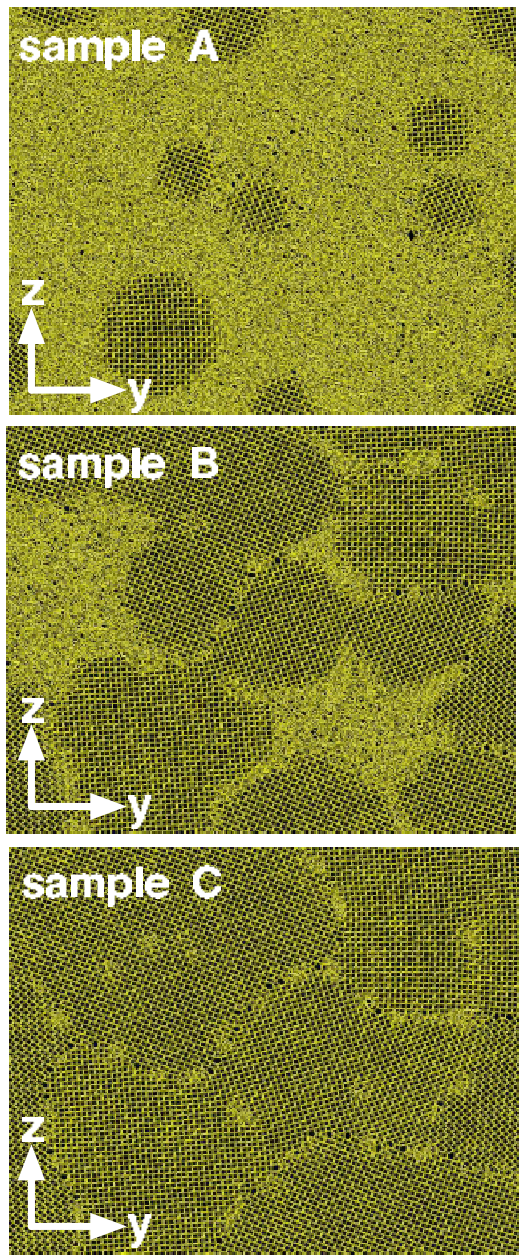}
\caption{(Color online). Snapshots of three nc-Si samples: $y-z$ planar projection after 0.05 ns (sample A), 1.0 ns (sample B), and 3.0 ns (sample C) annealing time at 1200 K.}
\label{snapshot}
\end{figure}

Because of the annealing, a spontaneous grain growth was observed during the simulations, thus mimicking the real recrystallization process. Three models of nc-Si (hereafter referred to as sample A, B and C) were obtained by saving the atomic coordinates after 0.05 ns, 1.0 ns and 3.0 ns of simulated recrystallization, respectively. They are shown in Fig.\ref{snapshot}, where just a portion of the simulation cell is reported for the sake of clarity. By visual inspection of such snapshot configurations, it is easy to extract a qualitative structural characterization. In sample A the nanocrystals are embedded into the amorphous phase and well separated (i.e. only amorphous/crystalline boundaries are present in the system). Sample B has a larger degree of structural order, due to grains of larger diameter. This system contains both a-c boundaries and grain boundaries. Finally, sample C corresponds to a fully recrystallized system, consisting of large crystalline domains with different crystallographic orientations separated by thin
grain boundaries. The three structures are representative of a wide class of nanocrystalline systems as for the a-c ratio and different boundary features.

The key computational tool used in our investigation is the concept of crystallinity, which was addressed to calculate accurately the a-c ratio, and so to identify the atoms forming the boundaries (where the role of hydrogen is especially relevant). To this aim we first calculated the structure factor (SF)
\begin{equation}
\Theta(t,T) = \frac{1}{N} \left| \sum_l e^{i\mathbf{k}\cdot \mathbf{r}_l}  \right|
\end{equation}
where ${\bf r}_l$ ($l=1,...,N$) are the atomic coordinates of the $N$ atoms in the system, $\mathbf{k}=(2\pi / d, 0, 0)$, and $d$ is the interplanar distance along the $[1,0,0]$ direction. The total crystallinity of any given sample was accordingly computed as
\begin{equation}
\chi_c(t)= \frac{\langle \Theta_{ac}(t,T)\rangle -\Theta_a}{\Theta_c(T)-\Theta_a}
\end{equation} 
where $\langle \Theta_{ac}(t,T)  \rangle$ is the average SF in the simulation cell, while $\Theta_c (T)$ and $\Theta_a$ are the SF's calculated for bulk c-Si and a-Si, respectively. We have found a constant $\Theta_a \sim 0.1$ over all investigated temperatures, while $\Theta_c(T)$ was calculated to decrease linearly with increasing temperature from $\Theta_c(T=0K)=1$ down to much smaller positive values. Nevertheless, we observed that $\Theta_c > \Theta_a$ for any $T$ below the c-Si melting temperature, so that  the c-Si and the a-Si regions in nanocrystalline samples can be easily identified by calculating the actual value of SF. In order to characterize the local structure within our samples, the simulation cell was divided into a mesh of $1\times50\times50$ subcells, each containing at least several hundreds of atoms. The local crystallinity values $\xi$ in each subcell have been interpolated as in 
Ref.[\onlinecite{ferraro-prb-nc}], thus obtaining a crystallinity map: it will be our main tool for the following quantitative analysis. The calculation of the local crystallinity value $\xi$ is useful also to identify the phase boundaries (both grain boundaries or a-c interfaces). By convention we set that if locally $\xi<0.3$ ($\xi>0.75$), then that given subcell is considered to contain amorphous (crystalline) matter. The number of remaining subcells (neither amorphous, nor crystalline) divided by the total number $1\times50\times50$ of subcells is a measure of the boundaries area, hereafter referred to as effective interface area. 

\begin{figure}
\includegraphics[width= 0.45\textwidth, angle=0]{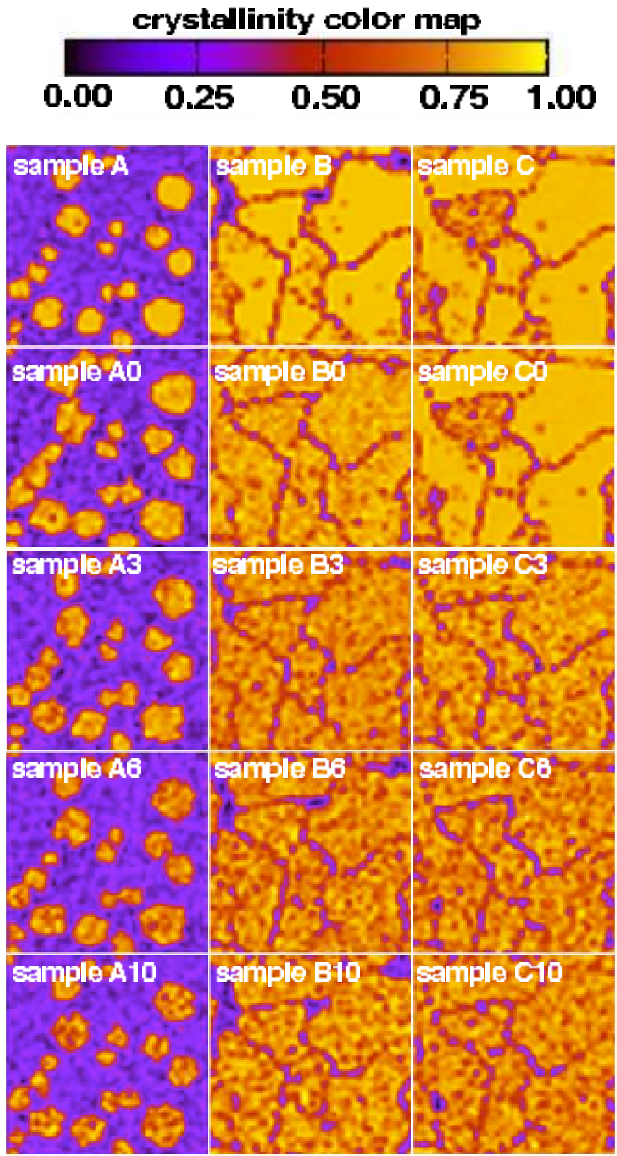}
\caption{(Color online) Local crystallinity maps of different $XN$ systems ($X$= sample A, B, or C; $N$= 0\%, 3\%, 6\%, or 10\% hydrogen content). Crystalline and amorphous regions are identified by yellow (light gray) and purple (dark gray) colors, respectively }
\label{cristall}
\end{figure}

In the first row of Fig.\ref{cristall} we report the crystallinity maps of samples A (left), B (middle), and C (right). Overall, the resulting picture is quite consistent with the previous qualitative structural characterization and it actually provides a quantitative estimation of the different a-c ratios in the three samples; in fact, they correspond to a total crystallinity of 20\%, 50\%, and $\sim$100\%, respectively.

In order to study the role of hydrogenation on the microstructure evolution of nc-Si, H atoms were randomly inserted in samples A, B and C at 3 different atomic concentrations (namely: $3\%$, $6\%$, and $10\%$), so generating nine different models of nc-Si:H, hereafter labeled as $XN$, where $X$ identifies the sample (i.e. $X$=A, B and C) and $N$ labels the H concentration (i.e. $N$=3\%, 6\%, and 10\%). 
The Si-H interactions were described  by  the Hansen and Vogl model potential,\cite{Hansenvogl} namely a Tersoff-like  potential\cite{Tersoff} including an extra environment-dependence in the angular energy term. This potential accurately describes the geometry and the formation energetics of interstitial H in c-Si.\cite{inanaga,hill} In order to further establish the reliability of the adopted force-field, we investigated the mobility of an isolated hydrogen atom in bulk c-Si and in a-Si, separately. In particular we evaluated its diffusivity $D(T)$ by calculating the corresponding mean square displacement and we obtained an activation energy for H migration of 0.44 eV (1.34 eV) in c-Si (a-Si). Both values compare well with literature values, respectively $\sim$0.5 eV and $\sim$1.5 eV (see Refs.[\onlinecite{panzarini, pantelides}] and references therein).

All samples were at first hydrogenated (see above), then relaxed by damped dynamics to get fully relaxed configurations, and eventually annealed at $T=1500$ K for $0.3$ ns in order to observe the microstructure evolution features. At this temperature (that is smaller than the melting of the crystalline phase) the thermal mobility of the H atoms in Si is high and the crystallization kinetics is fast. These properties are beneficial in order to reduce the simulation time and so the resulting computational workload. The same annealing procedure was also applied to the H-free samples, so as to directly compare with a microstructure evolution not affected by hydrogenation. The corresponding annealed samples have been labeled as A0, B0, and C0 consistently with the adopted notation.

The crystallinity maps calculated before and after the thermal annealing of both pure and hydrogenated samples are reported in Fig.\ref{cristall}, where each row corresponds to different H content and each column corresponds to a different a-c ratio of the initial system.
We consider first the A system (left column), namely the system with small a-c ratio. After the annealing and in absence of hydrogen (sample A0) the recrystallization is fast and the grains grow until some of them are in contact. Conversely, the same grains remain well separated when hydrogen is present, as clearly shown in samples A3, A6, and A10. At the largest H concentration the grains are practically unchanged with respect to the initial configuration. We conclude that the hydrogen is  detrimental for the  crystallization, particularly when the concentration is high. Since the annealing time is always the same, the above findings provide evidence that a slower recrystallization kinetics is found when H contamination is larger.

Similar conclusions can be drawn by analizing the central column of Fig.\ref{cristall}, where the microstructure evolution of sample B is reported. This system has a larger initial value of crystallinity, but there is still a sizeable amount of amorphous matter. After the annealing, the amorphous part of the H-free sample B0 has been almost fully recrystallized and only grain boundaries are present. Conversely, in the presence of H, the recrystallization is slower and this is proved by the fact that  some amorphous spots are still present after the same thermal annealing. Such effect is larger for larger H content (compare samples B10 and B).

The role of H is important also in the case of sample C (left column of Fig.\ref{cristall}), containing very little amorphous matter. In this case, H only affects the thickness of the grain boundaries, namely: the higher is the H content, the larger is the effective interface area. This result can be understood by considering the high mobility and the small solubility of hydrogen in c-Si: the H atoms initially located in the crystalline seeds rapidly diffuse\cite{panzarini} towards the interfaces, where they eventually get trapped. The H content  increases at the interface during the annealing, while an increase of the atomic scale disorder and the thickness of the grain boundaries is found as well (see C10).

\begin{figure}
\begin{center}
\includegraphics[width= 0.48\textwidth, angle=0]{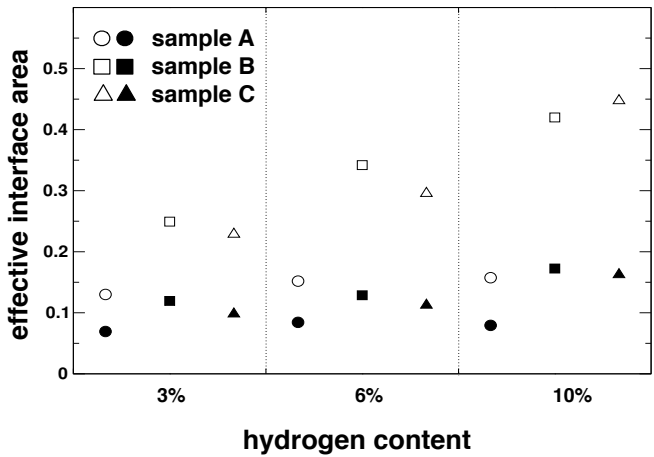}
\caption{Effective interface area versus hydrogen content before (filled symbols) and after (empty symbols) annealing.}
\label{interface}
\end{center}
\end{figure}

This effect is demonstrated quantitatively in Fig.\ref{interface} where the effective interface area as a function of the H concentration is reported before (filled symbols) and after (empty symbols) the annealing for the sample A (circles), B (squares) and C (triangles). In any case we observe that the higher is the H occurence the larger is the increment of the interface 
area after annealing. We finally observe that for a given value of  H content, the interface area has a larger increment in samples B and C than in sample A. This can be explained since they have a larger crystalline fraction and, therefore, a larger amount of H that can diffuse easily to the boundaries.
 
Finally, the analysis of the crystallinity map shows another important  phenomenon associated to the presence of H, namely the formation of Si$_n$H$_m$ aggregates. Let us consider the sample A and focus on the largest crystalline region (bottom right corner of panel labelled A of Fig.\ref{cristall}). Upon annealing, the crystallinity of this grain is practically unaffected, provided that no hydrogen is present, as clearly shown in panel A0 of the same figure. At variance, the crystallinity value remarkably decreases when H is present and the larger is the hydrogen amount, the larger is such an effect. The amorphous spots appearing with the grain correspond to a sizeable local reduction of the crystalline order due to the formation of Si$_n$H$_m$ aggregates. By visual inspection of the snapshot configurations (here not shown) we concluded that such complexes involving H atoms are found in both crystalline and amorphous phases.

In conclusion, by means of accurate computer experiments we provided evidence that hydrogen prevents full recrystallization of mixed-phase amorphous/crystalline silicon systems. This effect is enhanced by inceasing the H content. In particular, for nc-Si samples the presence of hydrogen impurities directly affects the thickness of the grain boundaries, by increasing their effective interface area. Finally, the formation of Si$_n$H$_m$ aggregates was reported, futher reducing the overall crystallinity.

\bibliography{manuscript-PRB}
\end{document}